\documentclass[letterpaper, 12pt]{article}
\usepackage[english]{babel}
\usepackage{amsfonts,amsmath,amssymb,bbm,amsthm,latexsym,amscd,mathrsfs}
\usepackage{ifthen,cite}
\usepackage[bookmarksnumbered=true]{hyperref}
\usepackage{times,fancyhdr}

\usepackage{amsmath}
\numberwithin{equation}{section}

\usepackage{graphicx}

\hypersetup{pdfpagetransition={Split}}

\newcommand{\evenhead}{Author \ name}
\newcommand{\oddhead}{Article \ name}
\newcommand{\theArticleName}{Article \ name}

\newcommand{\FirstPageHeading}[1]{\thispagestyle{empty}%
\noindent\raisebox{0pt}[0pt][0pt]{\makebox[\textwidth]{\protect\footnotesize \sf}}\par}

\newcommand{\ArticleName}[1]{\renewcommand{\theArticleName}{#1}\vspace{-2mm}\par\noindent {\LARGE\bf  #1\par}}
\newcommand{\Author}[1]{\vspace{5mm}\par\noindent {\Large  #1\par} \par\vspace{2mm}\par}
\newcommand{\Address}[1]{\vspace{2mm}\par\noindent {\it #1} \par}
\newcommand{\Email}[1]{\ifthenelse{\equal{#1}{}}{}{\par\noindent {\rm E-mail: }{\it  #1} \par}}
\newcommand{\URLaddress}[1]{\ifthenelse{\equal{#1}{}}{}{\par\noindent {\rm URL: }{\tt  #1} \par}}
\newcommand{\EmailD}[1]{\ifthenelse{\equal{#1}{}}{}{\par\noindent {$\phantom{\dag}$~\rm E-mail: }{\it  #1} \par}}
\newcommand{\URLaddressD}[1]{\ifthenelse{\equal{#1}{}}{}{\par\noindent {$\phantom{\dag}$~\rm URL: }{\tt  #1} \par}}

\newcommand{\Abstract}[1]{\vspace{6mm}\par\noindent\hspace*{10mm}
\parbox{140mm}{\small {\bf Abstract.} #1}\par}
\newcommand{\Keywords}[1]{\vspace{3mm}\par\noindent\hspace*{10mm}
\parbox{140mm}{\small {\bf Key words:} \rm #1}\par}
\newcommand{\Classification}[1]{\vspace{3mm}\par\noindent\hspace*{10mm}
\parbox{140mm}{\small {\it 2020 Mathematics Subject Classification:} \rm #1}\vspace{3mm}\par}
\newcommand{\ShortArticleName}[1]{\renewcommand{\oddhead}{#1}}
\newcommand{\AuthorNameForHeading}[1]{\renewcommand{\evenhead}{#1}}
\long\def\@makecaption#1#2{
  \sbox\@tempboxa{\small \textbf{#1.}\ \ #2}%
  \ifdim \wd\@tempboxa >\hsize
    {\small \textbf{#1.}\ \ #2}\par \else
    \global \@minipagefalse
    \hb@xt@\hsize{\hfil\box\@tempboxa\hfil}%
  \fi \vskip\belowcaptionskip}

\def\numberwithin#1#2{\@ifundefined{c@#1}{\@nocounterr{#1}}{%
  \@ifundefined{c@#2}{\@nocnterr{#2}}{%
  \@addtoreset{#1}{#2}%
  \toks@\@xp\@xp\@xp{\csname the#1\endcsname}%
  \@xp\xdef\csname the#1\endcsname
    {\@xp\@nx\csname the#2\endcsname.\the\toks@}}}}
\def\E^#1{{\buildrel #1 \over\vee}}
\newtheorem{theorem}{Theorem}
\newtheorem{criterion}{Criterion}

{\theoremstyle{definition}

}

\usepackage{tikz}

\begin{document}

\FirstPageHeading{V.I. Gerasimenko}

\ShortArticleName{Cumulant expansions}

\AuthorNameForHeading{V.I. Gerasimenko}

\ArticleName{\textcolor{blue!50!black}{Cumulant expansions of operator groups \\ of quantum many-particle systems}}

\Author{V.I. Gerasimenko$^\ast$\footnote{E-mail: \emph{gerasym@imath.kiev.ua}} and
I.V. Gapyak$^{\ast,}$$^\ast$$^\ast$\footnote{E-mail: \emph{ihapiak@knu.ua}}}

\Address{$^\ast$\hspace*{1mm}Institute of Mathematics of NAS of Ukra\"{\i}ne,\\
    \hspace*{3mm}3, Tereshchenkivs'ka Str.,\\
    \hspace*{3mm}01601, Ky\"{\i}v-4, Ukra\"{\i}ne}

\Address{$^\ast$$^\ast$Taras Shevchenko National University of Ky\"{\i}v,\\
    \hspace*{3mm}Department of Mechanics and Mathematics,\\
    \hspace*{3mm}2, Academician Glushkov Av.,\\
    \hspace*{3mm}03187, Ky\"{\i}v, Ukra\"{\i}ne}

\bigskip

\Abstract{The article presents a method of cluster expansions for groups of operators associated
with the von Neumann equations for states and the Heisenberg equations for observables, aiming to construct
generating operators for nonperturbative solutions to the Cauchy problem for hierarchies of evolution
equations of many-particle quantum systems.
}

\bigskip


\vspace{2pc}
\Classification{82C05, 82C40, 35Q20, 46N55, 47J35}

\makeatletter
\renewcommand{\@evenhead}{
\hspace*{-3pt}\raisebox{-7pt}[\headheight][0pt]{\vbox{\hbox to \textwidth {\thepage \hfil \evenhead}\vskip4pt \hrule}}}
\renewcommand{\@oddhead}{
\hspace*{-3pt}\raisebox{-7pt}[\headheight][0pt]{\vbox{\hbox to \textwidth {\oddhead \hfil \thepage}\vskip4pt\hrule}}}
\renewcommand{\@evenfoot}{}
\renewcommand{\@oddfoot}{}
\makeatother

\newpage
\vphantom{math}

\protect\textcolor{blue!50!black}{\tableofcontents}

\vspace{0.8cm}

\textcolor{blue!50!black}{\section{Introduction}}
Recent advances have been made in developing mathematical methods to describe the evolution of quantum many-particle systems.
This progress was mainly related to the challenge of deriving quantum kinetic equations as the scaling asymptotics of solutions
to hierarchies of fundamental evolution equations \cite{BPS},\cite{Ger12},\cite{P2021}. Additionally, modern techniques for
analyzing the propagation of correlations in complex information systems have contributed to this progress  \cite{SW},\cite{MW}.

As is well-known \cite{CGP97},\cite{Ger12}, many-particle systems are described in terms of a state and observables.
The mean-value functional of the observables defines the duality between the observables and the state. Consequently,
there are two equivalent approaches to describing the evolution using fundamental evolution equations for the observables
and the state, respectively. The conventional approach to describing the evolution of both finite and infinite numbers of
quantum particles is based on the concept of a state, specifically using reduced density operators. These operators are
governed by the BBGKY (Bogolyubov--Born--Green--Kirkwood--Yvon) hierarchy \cite{BogLect}.

In modern rigorous works, the solution of the BBGKY hierarchy for quantum many-particle systems is historically
represented by a perturbation theory series \cite{BogLect},\cite{Pe71}. Such a representation of the solution is used to
construct its scaling asymptotics, in particular, the mean-field asymptotics, which is described by the Vlasov quantum kinetic
equation \cite{BSh17},\cite{G13},\cite{PP09} and the Gross--Pitaevsky equation for the Bose condensate \cite{BSS22},\cite{EShY10},
or the weak-coupling asymptotics to derive the quantum Boltzmann kinetic equation \cite{BCEP3},\cite{CG}. Note that the use of
such a representation for the solution, due to the technical conditions associated with its existence, restricts the class of
interaction potentials and initial states of quantum particles, suitable from a physical point of view.

The article further examines the structure of series expansions that represent nonperturbative solutions to the Cauchy problem
for hierarchies of fundamental evolution equations related to both state and observable quantum many-particle systems. A method
for cluster expansions of groups of operators is developed, which allows for determining the cumulant expansions of the generating
operators for these solutions.

\textcolor{blue!50!black}{\section{Cumulants of operator groups of quantum systems}}
To identify the generating operators for the expansions of solutions to the Cauchy problem related to hierarchies of evolution
equations for quantum systems with many particles, we will introduce groups of operators associated with quantum systems of a
finite number of particles, along with their cumulant expansions.

In the following, we consider quantum systems of a non-fixed number of identical spinless particles, i.e., an arbitrary but finite
average number of particles that obey the Maxwell--Boltzmann statistics, and use a system of units where $h={2\pi\hbar}=1$ is Planck's
constant and $m =1$ is the particle mass. Let $\mathcal{H}_n=\mathcal{H}^{\otimes n}$ be an $n$-particle Hilbert space. The Fock space
over the space $\mathcal{H}$ is denoted by $\mathcal{F}_{\mathcal{H}}={\bigoplus\limits}_{n=0}^{\infty}\mathcal{H}_{n}$.

Let $\mathfrak{L}(\mathcal{H}_{n})$ be the space of bounded self-adjoint operators $b_{n}\in\mathfrak{L}(\mathcal{H}_{n})$ with operator
norm $\|.\|_{\mathfrak{L}(\mathcal{H}_{n})}$.
The self-adjoint operator $b_{n}$ is defined on the space $\mathcal{H}_{n}=\mathcal{H}^{\otimes n}$, hereinafter also denoted by the symbol
$b_{n}(1,\ldots,n)$, which for arbitrary values of the indices $(i_{1},\ldots,i_{n})\in(1,\ldots,n)$ satisfies the symmetry condition: $b_{n}(1,\ldots,n)=b_{n}(i_{1},\ldots,i_{n})$.
Let us introduce the space $\mathfrak{L}_\gamma(\mathcal{F}_{\mathcal{H}})$ of sequences of bounded operators
$b=(b_0,b_1(1),\ldots,b_n(1,\ldots,n),\ldots)$ with norm
\begin{eqnarray*}
   &&\|b\|_{\mathfrak{L}_\gamma(\mathcal{F}_{\mathcal{H}})}=
      \max_{n\geq0}\frac{\gamma^n}{n!}\,\|b_{n}\|_{\mathfrak{L}(\mathcal{H}_{n})},
\end{eqnarray*}
where $0<\gamma<1$ is a real number. The subspace of finite sequences of degenerate operators whose kernels are infinitely differentiable
functions with compact supports is denoted by $\mathfrak{L}_0(\mathcal{F}_{\mathcal{H}})$.

Accordingly, let $\mathfrak{L}^{1}(\mathcal{H}_{n})$ be the space of nuclear operators
$f_{n}\equiv f_{n}(1,\ldots,n)\in\mathfrak{L}^{1}(\mathcal{H}_{n})$, which for arbitrary values of the indices
$(i_{1},\ldots,i_{n})\in(1,\ldots,n)$ satisfy the symmetry condition: $f_{n}(1,\ldots,n)=f_{n}(i_{1},\ldots,i_{n})$, with norm:
$\|f_{n}\|_{\mathfrak{L}^{1}(\mathcal{H}_{n})}=\mathrm{Tr}_{1,\ldots,n}|f_{n}(1,\ldots,n)|,$ where the symbol $\mathrm{Tr}_{1,\ldots,n}$
denote the partial traces of the operator $f_{n}$. Let us introduce the space $\mathfrak{L}^{1}_\alpha(\mathcal{F}_{\mathcal{H}})$ of
sequences of trace-class operators $f=(f_0,f_1(1),\ldots,f_n(1,\ldots,n),\ldots)$ with norm
\begin{eqnarray*}
   &&\|f\|_{\mathfrak{L}^{1}_\alpha(\mathcal{F}_{\mathcal{H}}})=
      \sum_{n=0}^\infty \alpha^n \mathrm{Tr}_{1,\ldots,n}|f_{n}(1,\ldots,n)|,
\end{eqnarray*}
where $\alpha>1$ is a real number. The subspace of finite sequences of degenerate operators whose kernels are infinitely
differentiable functions with compact supports will be denoted by $\mathfrak{L}^{1}_0(\mathcal{F}_{\mathcal{H}})$.

In the space $\mathfrak{L}_\gamma(\mathcal{F}_{\mathcal{H}})$ for arbitrary values of $t\in\mathbb{R}$, a one-parameter
family of mappings $\mathcal{G}(t)=\oplus_{n=0}^\infty\mathcal{G}_n(t)$ is defined, where the mapping $\mathcal{G}_n(t)$
in the subspace $\mathfrak{L}(\mathcal{H}_{n})$ is defined by the following formula
\begin{eqnarray}\label{grG}
&&\mathbb{R}^1\ni t\mapsto\mathcal{G}_n(t)b_n\doteq e^{itH_{n}}b_n e^{-itH_{n}},
\end{eqnarray}
and the self-adjoint operator
$H_{n}={\sum\limits}_{j=1}^{n}K(j)+{\sum\limits}_{j_{1}<j_{2}=1}^{n}\Phi(j_{1},j_{2})$ is the Hamiltonian of the system
of $n$ particles, i.e., the operator $K(j)$ is the kinetic energy operator of $j$ particle, $\Phi$ is the bounded pair
interaction potential operator.
In the space $\mathfrak{L}(\mathcal{H}_{n})$ the one-parameter mapping (\ref{grG}) forms a $\ast$-weakly continuous group
of operators, the infinitesimal generator of which, on the domain of its definition, coincides with the Heisenberg operator
\begin{eqnarray}\label{info}
    &&\mathcal{N}_n b_n\doteq-i\,(b_n H_n - H_n b_n).
\end{eqnarray}

Let us introduce the concept of cumulants (semi-invariants) of operator groups (\ref{grG}), that is, the connected part of
operator groups (\ref{grG}).

The cumulant of the $s$th order of groups of operators (\ref{grG}) is defined by the following expansions \cite{Ger12}:
\begin{equation}\label{cum}
   \mathfrak{A}_{s}(t,1,\ldots,s)\doteq
      \sum\limits_{\mbox{\scriptsize $\begin{array}{c}\mathrm{P}:(1,\ldots,s)=
       \bigcup_{i}X_{i}\end{array}$}}(-1)^{|\mathrm{P}|-1}(|\mathrm{P}|-1)!\,
       \prod_{X_i\subset \mathrm{P}}\mathcal{G}_{|X_i|}(t,X_i),
\end{equation}
where the symbol $\sum_\mathrm{P}$ denotes the sum over all partitions of $\mathrm{P}$ of the set of indices $(1,\ldots,s)$
into $|\mathrm{P}|$ nonempty subsets $X_i\in(1,\ldots,s)$ that do not intersect. Let us give examples of cumulant expansions
\eqref{cum} of groups of operators (\ref{grG}):
\begin{eqnarray*}
    &&\mathfrak{A}_{1}(t,1)=\mathcal{G}_{1}(t,1), \\
    &&\mathfrak{A}_{2}(t,1,2)=\mathcal{G}_{2}(t,1,2)-\mathcal{G}_{1}(t,1)\mathcal{G}_{1}(t,2),\\
    &&\mathfrak{A}_{3}(t,1,2,3)=\mathcal{G}_{3}(t,1,2,3)-\mathcal{G}_{1}(t,1)\mathcal{G}_{2}(t,2,3)-
    \mathcal{G}_{1}(t,2)\mathcal{G}_{2}(t,1,3)-\\
    &&\hskip+7mm \mathcal{G}_{1}(t,3)\mathcal{G}_{2}(t,1,2)+
    2!\,\mathcal{G}_{1}(t,1)\mathcal{G}_{1}(t,2)\mathcal{G}_{1}(t,3),\\
   &&\hskip+7mm \vdots
\end{eqnarray*}

Note that the cumulants of operator groups \eqref{cum} are solutions of recursive equations known as cluster expansions
of operator groups(\ref{grG})
\begin{equation}\label{exp}
    \mathcal{G}_{n}(t,1,\ldots,n)=
      \sum\limits_{\mbox{\scriptsize $\begin{array}{c}\mathrm{P}:(1,\ldots,n)=
       \bigcup_{i}X_{i}\end{array}$}}\prod_{X_i\subset \mathrm{P}}\mathfrak{A}_{|X_i|}(t,X_i),\quad n\geq 1,
\end{equation}
where the notation of the formula \eqref{cum} is used. The structure of cluster distributions is determined by the structure
of the infinitesimal generator \eqref{info} of the group of operators (\ref{grG}).

Let us present some characteristic properties of cumulants of groups of operators \eqref{cum}.

In the case of a quantum system of non-interacting particles, we have
\begin{eqnarray*}
   &&\mathfrak{A}_{s}(t,1,\ldots,s)=0,\quad s\geq2.
\end{eqnarray*}
Indeed, since in this case the equality $\mathcal{G}_{n}(t,1,\ldots,n)=\prod_{j=1}^{n}\mathcal{G}_{1}(t,j),$ holds
for the group of operators (\ref{grG}), we establish
\begin{eqnarray*}
  &&\mathfrak{A}_{s}(t,1,\ldots,s)=\sum\limits_{\mathrm{P}:\,(1,\ldots,s)={\bigcup}_i X_i}
     (-1)^{|\mathrm{P}|-1}(|\mathrm{P}|-1)!\prod\limits_{X_{i}\subset
     \mathrm{P}}\prod_{j_i=1}^{|X_{i}|}\mathcal{G}_{1}(t,j_i)=\\
  &&=\sum\limits_{k=1}^{s}(-1)^{k-1}\mathrm{s}(s,k)(k-1)!\prod_{j=1}^{s}\mathcal{G}_{1}(t,j)=0,
\end{eqnarray*}
where $\mathrm{s}(s,k)$ are Stirling numbers of the second kind, and the equality is used
\begin{eqnarray*}\label{Stirl}
  &&\hskip-5mm \sum\limits_{\mathrm{P}:\,(1,\ldots,s)={\bigcup}_i X_i}
    (-1)^{| \mathrm{P}|-1}(|\mathrm{P}|-1)!=
    \sum\limits_{k=1}^{s}(-1)^{k-1}\mathrm{s}(s,k)(k-1)!=\delta_{s,1},
\end{eqnarray*}
where $\delta_{s,1}$ is the Kronecker symbol.

If $b_{s}\in\mathfrak{L}(\mathcal{H}_{s})$, then for the cumulant of $s$ order \eqref{cum}, the following estimate
is valid
\begin{eqnarray}\label{est}
   &&\big\|\mathfrak{A}_{s}(t,1,\ldots,s)b_{s}\big\|_{\mathfrak{L}(\mathcal{H}_{s})}
      \leq\sum\limits_{\mathrm{P}:\,(1,\ldots,s)={\bigcup}_i X_i}
      (|\mathrm{P}|-1)!\,\big\|b_{s}\big\|_{\mathfrak{L}(\mathcal{H}_{s})}\leq \\
   &&\leq\sum\limits_{k=1}^{s}\mathrm{s}(s,k)(k-1)!\big\|b_{s}\big\|_{\mathfrak{L}(\mathcal{H}_{s})}
      \leq s!\,e^{s}\,\big\|b_{s}\big\|_{\mathfrak{L}(\mathcal{H}_{s})},\nonumber
\end{eqnarray}
where $\mathrm{s}(s,k)$ are Stirling numbers of the second kind.

The functional of the mean values (mathematical expectation) of sequences from the space
$\mathfrak{L}_\gamma(\mathcal{F}_{\mathcal{H}})$ is defined by a continuous linear functional, represented
by the following series expansion \cite{Ger12}:
\begin{eqnarray}\label{averageD}
     &&\langle b\rangle=(I,f)^{-1}(b,f)\doteq(I,f)^{-1}\sum\limits_{n=0}^{\infty}\frac{1}{n!}
         \,\mathrm{Tr}_{1,\ldots,n}\,b_{n}\,f_{n},
\end{eqnarray}
where the sequence $f\in \mathfrak{L}^1_\alpha(\mathcal{F}_{\mathcal{H}})$ and the factor
$(I,f)={\sum\limits}_{n=0}^{\infty}\frac{1}{n!}\mathrm{Tr}_{1,\ldots,n}f_{n}$ is the normalizing coefficient.
If $b\in\mathfrak{L}_\gamma(\mathcal{F}_{\mathcal{H}})$ and $f\in\mathfrak{L}^1_\alpha(\mathcal{F}_{\mathcal{H}})$
the functional \eqref{averageD} exists.

In the space $\mathfrak{L}^1_\alpha(\mathcal{F}_{\mathcal{H}})$, the dual in the sense of the functional (\ref{averageD})
one-parameter mapping $\mathcal{G}^\ast(t)=\oplus_{n=0}^\infty \mathcal{G}^\ast_n(t)$, conjugate to the group (\ref{grG}),
is defined as
\begin{eqnarray}\label{grGs}
    &&\mathbb{R}^1\ni t\mapsto\mathcal{G}^{\ast}_n(t)f_n\doteq e^{-itH_{n}}f_n e^{itH_{n}},
\end{eqnarray}
and forms a strongly continuous isometric group of operators that preserves positivity and self-adjointness of operators
\cite{CM21}. The infinitesimal generator of the group (\ref{grGs}) on the subspace $\mathfrak{L}^{1}_0(\mathcal{F}_{\mathcal{H}})$
coincides with the von Neumann operator
\begin{eqnarray}\label{infd}
    &&\mathcal{N}^{\ast}_n f_n\doteq-i\,(H_n f_n - f_n H_n).
\end{eqnarray}

The cumulant of the $s$th order of operator groups (\ref{grGs}) is defined by the following cumulant expansion \cite{Ger12}:
\begin{equation}\label{cumd}
   \mathfrak{A}_{s}^\ast(t,1,\ldots,s)\doteq
      \sum\limits_{\mbox{\scriptsize $\begin{array}{c}\mathrm{P}:(1,\ldots,s)=
       \bigcup_{i}X_{i}\end{array}$}}(-1)^{|\mathrm{P}|-1}(|\mathrm{P}|-1)!\,
       \prod_{X_i\subset \mathrm{P}}\mathcal{G}^\ast_{|X_i|}(t,X_i),
\end{equation}
where the symbol $\sum_\mathrm{P}$ denotes the sum over all partitions of $\mathrm{P}$ of the set of indices $(1,\ldots,s)$
into $|\mathrm{P}|$ nonempty subsets $X_i\in(1,\ldots,s)$ that do not intersect.
Note that the cumulants of operator groups \eqref{cumd} are solutions of recursive equations that have the structure of
cluster expansions of operator groups (\ref{grGs})
\begin{equation}\label{expd}
    \mathcal{G}^\ast_{n}(t,1,\ldots,n)=
      \sum\limits_{\mbox{\scriptsize $\begin{array}{c}\mathrm{P}:(1,\ldots,n)=
       \bigcup_{i}X_{i}\end{array}$}}\prod_{X_i\subset \mathrm{P}}\mathfrak{A}^\ast_{|X_i|}(t,X_i), \quad n\geq 1,
\end{equation}
where the notation of the formula \eqref{cumd} is used. In the following, a generalization of cluster and cumulant expansions
of operator groups for clusters of quantum particles and particles will be introduced.

\textcolor{blue!50!black}{\section{Hierarchies of evolution equations of quantum systems}}
Quantum systems of a non-fixed number of identical particles are described in terms of a sequence of self-adjoint operators of
observables $A=(A_0,A_1,\ldots,A_n,\ldots)\in\mathfrak{L}_\gamma(\mathcal{F}_{\mathcal{H}})$ and a sequence of positive density
operators $D=(D_0,D_1,\ldots,D_n,\ldots)\in\mathfrak{L}^1_\alpha(\mathcal{F}_{\mathcal{H}})$, which describes the state of the
particle system \cite{Ger12}. The groups of operators (\ref{grG}) and (\ref{grGs}) determine the evolution of the observables
and the state of quantum systems of many particles, respectively.

If $A(0)\in\mathfrak{L}_\gamma(\mathcal{F}_{\mathcal{H}})$, then for $t\in\mathbb{R}$ the unique solution of the Cauchy problem
for the sequence of Heisenberg equations
\begin{eqnarray}\label{A(t)}
&&\frac{d}{dt}A(t)=\mathcal{N}A(t),
\end{eqnarray}
where the generator $\mathcal{N}=\oplus_{n=0}^\infty\mathcal{N}_n$ is given by the formula \eqref{info}, represented by the group
of operators (\ref{grG})
\begin{eqnarray}\label{gA(t)}
&&A(t)=\mathcal{G}(t)A(0)\doteq\oplus_{n=0}^\infty\,\mathcal{G}_n(t)A_n(0).
\end{eqnarray}

For the mean value functional (\ref{averageD}) of a sequence of observables (\ref{gA(t)}), the following equivalent representation
is valid
\begin{eqnarray*}
&&\langle A(t)\rangle=(I,D(0))^{-1}(A(t),D(0))=(I,D(t))^{-1}(A(0),D(t)),
\end{eqnarray*}
where the sequence of density operators $D(t)\in\mathfrak{L}^1_\alpha(\mathcal{F}_{\mathcal{H}})$ is represented by the group of
operators (\ref{grGs})
\begin{eqnarray}\label{gD(t)} &&D(t)=\mathcal{G}^\ast(t)D(0)\doteq\oplus_{n=0}^\infty\,\mathcal{G}^\ast_n(t)D_n(0).
\end{eqnarray}
For $t\in\mathbb{R}$ the sequence (\ref{gD(t)}) is the only solution to the Cauchy problem for the sequence of von Neumann equations
\begin{eqnarray}\label{D(t)}
&&\frac{d}{dt}D(t)=\mathcal{N}^\ast D(t),
\end{eqnarray}
whose generator $\mathcal{N}^\ast=\oplus_{n=0}^\infty\mathcal{N}^\ast_n$ is given by the formula \eqref{infd}.

As is known \cite{BogLect},\cite{Ger12}, there is another method of describing the states and observables of many-particle quantum
systems, which is based on an equivalent representation of the mean value functional (\ref{averageD}) observables in terms of a
sequence of reduced observables $B(t)\in\mathfrak{L}_\gamma(\mathcal{F}_{\mathcal{H}})$ and reduced density operators
$F(0)\in\mathfrak{L}^1_\alpha(\mathcal{F}_{\mathcal{H}})$, namely,
\begin{eqnarray}\label{B(t)}
&&\langle A(t)\rangle=(I,D(0))^{-1}(A(t),D(0))=(B(t),F(0)).
\end{eqnarray}
Note that the ability to describe the observables and the state using the corresponding reduced operators is the result of
summarizing the series in the expression \eqref{averageD}, taking into account the normalizing factor.

To construct such a representation of the mean value functional of the bounded operators observed on sequences, we introduce
a bounded operator in the space $\mathfrak{L}_\gamma(\mathcal{F}_{\mathcal{H}})$ analogous to the creation operator
\begin{eqnarray}\label{a+}
&&\hskip-5mm(\mathfrak{a}^{+}b)_{n}(1,\ldots,n)\doteq
\sum_{j=1}^n\,b_{n-1}((1,\ldots,n)\setminus j)\otimes\mathbbm{1}_{(j)}, \quad n\geq 1,
\end{eqnarray}
where the symbol $\mathbbm{1}_{(j)}$ denotes the unit operator in the $j$ Hilbert subspace of the space $\mathcal{H}^{\otimes n}$.
On sequences of nuclear operators from the space $\mathfrak{L}^1_\alpha(\mathcal{F}_{\mathcal{H}})$ we define a bounded adjoint
operator to the operator (\ref{a+}) in the sense of the functional (\ref{averageD}), which is an analogue of the annihilation
operator,
\begin{eqnarray}\label{a}
&&\hskip-5mm(\mathfrak{a}f)_{n}(1,\ldots,n)=\mathrm{Tr}_{n+1}\,f_{n+1}(1,\ldots,n,n+1),\quad n\geq 1.
\end{eqnarray}
Then, due to the validity of the equalities:
\begin{eqnarray*}
&&(b,f)=(e^{\mathfrak{a^{+}}}e^{-\mathfrak{a^{+}}}b,f)=(e^{-\mathfrak{a^{+}}}b,e^{\mathfrak{a}}f),
\end{eqnarray*}
for the mean-value functional of observables (\ref{averageD}) we derive representation \eqref{B(t)}, where
the sequence of reduced observables is given by the formula
\begin{eqnarray}\label{ro}
&& B(t)=e^{-\mathfrak{a^{+}}}A(t),
\end{eqnarray}
and the sequence of reduced density operators is given by the following expression
\begin{eqnarray}\label{rdf1}
&&F(0)=(I,D(0))^{-1}e^{\mathfrak{a}}D(0).
\end{eqnarray}

Thus, according to the definition of the operator $e^{-\mathfrak{a^{+}}}$, the sequence of reduced
observables (\ref{ro}) in component-wise form is represented by the following expansions:
\begin{eqnarray}\label{moo}
&&\hskip-5mm B_s(t,1,\ldots,s)=\\
&&\sum_{n=0}^s\,\frac{(-1)^n}{n!}\sum_{j_1\neq\ldots\neq j_{n}=1}^s
(\mathcal{G}(t)A(0))_{s-n}((1,\ldots,s)\setminus (j_1,\ldots,
j_{n}))\otimes\mathbbm{1}_{(j_1)}\otimes\ldots\otimes\mathbbm{1}_{(j_n)},\nonumber\\
&&\hskip-5mm s\geq1.\nonumber
\end{eqnarray}

We emphasize that for the mean value functional (\ref{B(t)}) of the sequence of reduced observables (\ref{ro}) the
equivalent representation in terms of the evolution of the state of the quantum system of particles
\begin{eqnarray}\label{F(t)}
&&(B(t),F(0))=(B(0),F(t)),
\end{eqnarray}
where according to the definition of the operator $e^{\mathfrak{a}}$, the sequence of reduced density operators
(\ref{rdf1}) in component-wise form is represented by the following expansions into a series:
\begin{eqnarray}\label{ms}
&&\hskip-5mm F_{s}(t,1,\ldots,s)\doteq (I,D(0))^{-1}\sum\limits_{n=0}^{\infty}\frac{1}{n!}\,
\mathrm{Tr}_{s+1,\ldots,s+n}\,(\mathcal{G}^\ast(t)D(0))_{s+n}(1,\ldots,s+n),\\
&&\hskip-5mm s\geq1.\nonumber
\end{eqnarray}

Note that the evolution of many-particle quantum systems is traditionally described by the BBGKY hierarchy for reduced
density operators (\ref{ms}), namely, the sequence (\ref{ms}) is determined by the Cauchy problem for the BBGKY hierarchy
\cite{BogLect},\cite{Ger12}:
\begin{eqnarray}\label{h}
&&\hskip-5mm\frac{d}{dt}F(t)=\mathcal{N}^\ast F(t)+\big[\mathfrak{a},\mathcal{N}^\ast\big]F(t),\\
\nonumber\\
\label{hi}
&&\hskip-5mmF(t)|_{t=0}=F(0),
\end{eqnarray}
where the symbol $\big[\cdot,\cdot \big]$ denotes the commutator of the operator (\ref{a}) and the von Neumann operator
(\ref{infd}) which is the generator of the group of operators (\ref{grGs}). Thus, the second term of the generator of the
hierarchy of evolution equations (\ref{h}) componentwise has the form:
\begin{eqnarray*}
&&\hskip-5mm (\big[\mathfrak{a},\mathcal{N}^\ast\big]f)_{s}(1,\ldots,s)=\sum_{j=1}^s\,\mathrm{Tr}_{n+1}\,
\mathcal{N}^\ast_{\mathrm{int}}(j,s+1)f_{s+1}(1,\ldots,s+1),\quad s\geq 1. \nonumber
\end{eqnarray*}
where the operator $\mathcal{N}^\ast_{\mathrm{int}}(j,s+1)f_{s+1}\doteq-i\,(\Phi(j,s+1)f_{s+1}-f_{s+1}\Phi(j,s+1))$
is determined by the particle interaction potential.

As noted, an equivalent approach to describing the evolution of quantum many-particle systems is based on
reduced observables (\ref{moo}), which are defined by the Cauchy problem for the hierarchy of evolution equations \cite{GB}:
\begin{eqnarray}\label{dh}
&&\hskip-5mm\frac{d}{dt}B(t)=\mathcal{N}B(t)+\big[\mathcal{N},\mathfrak{a}^+\big]B(t),\\
\nonumber\\
\label{dhi}
&&\hskip-5mm B(t)|_{t=0}=B(0),
\end{eqnarray}
where the Heisenberg operator $\mathcal{N}$ is the generator (\ref{info}) of the group of operators (\ref{grGs}),
the symbol $\big[ \cdot,\cdot \big]$ denotes the commutator of the operators, which componentwise has form:
\begin{eqnarray*}
&&(\big[\mathcal{N},\mathfrak{a}^+\big]b)_{s}(1,\ldots,s)
=\sum_{j_1\neq j_{2}=1}^s\mathcal{N}_{\mathrm{int}}(j_1,j_{2})b_{s-1}((1,\ldots,s)\setminus j_1)
\otimes\mathbbm{1}_{(j_1)},\quad s\geq 2. \nonumber
\end{eqnarray*}
In fact, the component-wise hierarchy of evolutionary equations (\ref{dh}) is a sequence of recurrent evolutionary
equations \cite{GB}.

The BBGKY hierarchy (\ref{h}) is dual to the hierarchy of evolution equations (\ref{dh}), i.e. the generator of the
BBGKY hierarchy for reduced density operators is the adjoint operator of the generator of the hierarchy of evolution
equations (\ref{dh}) for operators of reduced observables in the sense of the mean value functional \eqref{averageD}.
The structure of the generators of the specified hierarchy of evolution equations for quantum systems of particles
with an arbitrary $n$-ary interaction potential is established in \cite{Ger12}.

\textcolor{blue!50!black}{\section{Nonperturbative solution of the BBGKY hierarchy of quantum systems}}
The solution of the Cauchy problem for the BBGKY hierarchy for quantum many-particle systems is traditionally
represented in the form of a perturbation theory series (an iteration series with respect to the evolution of
the state of a group of particles). Such a representation of the solution is used to construct its scaling
asymptotics. This section formulates a method for constructing a solution of the Cauchy problem for the BBGKY
hierarchy that is not based on perturbation theory methods.

The nonperturbative solution of the Cauchy problem of the BBGKY hierarchy (\ref{h}),(\ref{hi}) for a sequence
of reduced density operators is represented by the following expansions into series \cite{GerS}:
\begin{eqnarray}\label{se}
&&\hskip-7mm F_s(t,1,\ldots,s)=\\
&&\hskip-5mm\sum_{n=0}^\infty\frac{1}{n!}\,\mathrm{Tr}_{s+1,\ldots,s+n}\,
\mathfrak{A}_{1+n}^\ast(t,\{1,\ldots,s\},s+1,\ldots,s+n)F_{s+n}(0,1,\ldots,s+n),\quad s\geq1,\nonumber
\end{eqnarray}
where the generating operators $\mathfrak{A}^\ast_{1+n}(t),\,n\geq0,$ of the series are represented by cumulative
expansions \eqref{cumd} of groups of operators (\ref{grGs}) of quantum particle systems. The cumulant of the $(1+n)$th
order of the groups of operators of a cluster of $s$ quantum particles and $n$ particles is given by the following expansion:
\begin{eqnarray}\label{cumulant}
&&\hskip-7mm \mathfrak{A}_{1+n}^\ast(t,\{1,\ldots,s\},s+1,\ldots,s+n)\doteq\\
&&\sum\limits_{\mathrm{P}:\,(\{1,\ldots,s\},\,s+1,\ldots,s+n)={\bigcup}_i X_i}
(-1)^{\mathrm{|P|}-1}({\mathrm{|P|}-1})!\prod_{X_i\subset\mathrm{P}}
\mathcal{G}^\ast_{|\theta(X_i)|}(t,\theta(X_i)),\nonumber
\end{eqnarray}
where the symbol $\{1,\ldots,s\}$ denotes the set consisting of one element of indices $(1,\ldots,s)$,
the declustering map $\theta$ is defined by the formula: $\theta(\{1,\ldots,s\})=(1,\ldots,s)$ and
the notation \eqref{cumd} introduced above is used.
Here are some examples of cumulant expansions \eqref{cumulant} of groups of operators (\ref{grGs})
of the von Neumann equations:
\begin{eqnarray*}
&&\hskip-7mm\mathfrak{A}^\ast_{1}(t,\{1,\ldots,s\})=\mathcal{G}^\ast_{s}(t,1,\ldots,s),\\
&&\hskip-7mm \mathfrak{A}^\ast_{1+1}(t,\{1,\ldots,s\},s+1)=
\mathcal{G}^\ast_{s+1}(t,1,\ldots,s+1)-\mathcal{G}^\ast_{s}(t,1,\ldots,s)\mathcal{G}^\ast_{1}(t,s+1),\\
&&\hskip-7mm\vdots
\end{eqnarray*}

Cumulants \eqref{cumulant} of groups of operators (\ref{grGs}) are defined as solutions of cluster expansions
\eqref{expd} of groups of operators (\ref{grGs}), namely the following recursive relations:
\begin{eqnarray}\label{cex}
&&\hskip-7mm \mathcal{G}_{s+n}^\ast(t,1,\ldots,s,s+1,\ldots,s+n)=\sum\limits_{\mathrm{P}:\,(\{1,\ldots,s\},\,s+1,\ldots,s+n)=
\bigcup_i X_i}\,\prod\limits_{X_i\subset\mathrm{P}}\mathfrak{A}^\ast_{|X_i|}(t,X_i),\quad n\geq 0,
\end{eqnarray}
where the symbol ${\sum}_{\mathrm{P}}$ means the sum over all partitions of $\mathrm{P}$ of the set
$(\{1,\ldots,s\},s+1,\ldots,s+n)$ into $|\mathrm{P}|$ non-empty subsets $X_i$ that do not intersect each other.

Since, similarly to the estimate \eqref{est} for the generating operator (\ref{cumulant}) of the series for reduced
density operators, we have
\begin{eqnarray*}
&&\hskip-5mm \big\|\mathfrak{A}^\ast_{1+n}(t)f_{s+n}\big\|_{\mathfrak{L}^{1}(\mathcal{H}_{s+n})}
\leq n!e^{n+2}\big\|f_{s+n}\big\|_{\mathfrak{L}^{1}(\mathcal{H}_{s+n})},\nonumber
\end{eqnarray*}
then under the condition $\alpha>e$ the series (\ref{se}) converges in the norm of the space
$\mathfrak{L}^{1}_{\alpha}(\mathcal{F}_{\mathcal{H}})=\oplus_{n=0}^\infty \alpha^n\mathfrak{L}^{1}(\mathcal{H}_{n})$
and the valid inequality
\begin{eqnarray*}\label{Fes}
&&\|F(t)\|_{\mathfrak{L}^{1}_{\alpha}(\mathcal{F}_{\mathcal{H}})}\leq
e^{2}(1-\frac{e}{\alpha})^{-1}\|F(0)\|_{\mathfrak{L}^{1}_{\alpha}(\mathcal{F}_{\mathcal{H}})}.
\end{eqnarray*}
The parameter $\alpha$ is interpreted as the inverse of the average number of particles.

The following criterion holds.
\begin{criterion} A sequence of reduced density operators represented by series expansions
(\ref{se}) is a solution of the Cauchy problem for the BBGKY hierarchy (\ref{h}),(\ref{hi}), if and only if
the generating operators (\ref{cumulant}) of the series (\ref{se}) are solutions of the cluster expansions (\ref{cex})
of the operator groups (\ref{grGs}) of the von Neumann equations.
\end{criterion}

A necessary condition means that the cluster expansions (\ref{cex}) of the operator groups (\ref{grGs}) determine the
structure of the generating operators of the solution of the Cauchy problem of the BBGKY hierarchy (\ref{h}),(\ref{hi}).
These recurrence relations are derived from the definition (\ref{ms}) of reduced density operators.

A sufficient condition means that the infinitesimal generator of the one-parameter mapping (\ref{se}) coincides
with the generator of the BBGKY hierarchy (\ref{h}).

Indeed, in the space $\mathfrak{L}^{1}_{\alpha}(\mathcal{F}_{\mathcal{H}})$ the following existence theorem is valid \cite{GerS}.
\begin{theorem} If $\alpha>e$, the nonperturbative solution of the Cauchy problem of the BBGKY hierarchy (\ref{h}),(\ref{hi})
is represented by series expansions (\ref{se}), whose generating operators are the cumulants of the corresponding order of the
groups of operators of the von Neumann equation (\ref{grGs}).

For initial states $F(0)\in \mathfrak{L}^{1}_{0}(\mathcal{F}_{\mathcal{H}})$ the sequence of reduced density operators (\ref{se})
is the unique global strong solution, and for arbitrary initial states $F(0)\in \mathfrak{L}^{1}_{\alpha}(\mathcal{F}_{\mathcal{H}})$
the unique global weak solution.
\end{theorem}

The proof of this theorem is analogous to the proof of the existence theorem for the von Neumann equations (\ref{D(t)}) of many
quantum particles in the space of sequences of nuclear operators \cite{Ger12},\cite{GerS}.

Note that for classical particle systems the structure of the first few terms of the series (\ref{se}) was established in
the works \cite{C62},\cite{G56} on the basis of an analogue of cluster expansions of reduced equilibrium distribution functions.

\textcolor{blue!50!black}{\section{Representation of reduced density operators by the perturbation theory}}
Cluster expansions of operator groups form the basis for the classification of all possible representations of the solutions
of the Cauchy problem for the BBGKY hierarchy (\ref{h}),(\ref{hi}) of quantum many-particle systems. In a special case, the
nonperturbative solution (\ref{se}) of the BBGKY hierarchy for quantum particle systems can be represented as a perturbation
theory series (an iteration series) \cite{BogLect},\cite{Pe71} as a result of applying analogs of the Duhamel equation to the
cumulants (\ref{cumulant}) of operator groups (\ref{grGs}). Traditionally, perturbation theory methods are used to study the
problem of rigorous derivation of quantum kinetic equations in the scaling limits of the solution of the BBGKY hierarchy.
For the first time, a rigorous justification of the solution in the form of a series of iterations in the space of sequences
of trace-class operators was obtained in the work of D. Ya. Petrina \cite{Pe71}.

Let us give some examples of solution representations to the Cauchy problem for the BBGKY hierarchy (\ref{h}),(\ref{hi}),
which, as special cases of the representation (\ref{se}), were established independently. For this purpose, we arrange
the terms in the expression of the cumulant expansion (\ref{cumulant}) in a new order with respect to the groups of
operators acting in the subspaces denoted by the indices $(1,\ldots,s)$, namely,
\begin{eqnarray}\label{peregr}
&&\hskip-7mm\mathfrak{A}^\ast_{1+n}(t,\{1,\ldots,s\},s+1,\ldots,s+n)=\\
&&\sum\limits_{Y\subset\,(s+1,\ldots,s+n)}\mathcal{G}^\ast_{s+|Y|}(t,(1,\ldots,s)\cup\,Y)
\sum\limits_{\mathrm{P}\,:(s+1,\ldots,s+n)\setminus Y={\bigcup\limits}_i Y_i}
(-1)^{|\mathrm{P}|}|\mathrm{P}|!\prod_{Y_i\subset\mathrm{P}}
\mathcal{G}^\ast_{|Y_{i}|}(t,Y_{i}).\nonumber
\end{eqnarray}

If $Y_{i}\subset(s+1,\ldots,s+n)$, then for the trace-class operators $F_{s+n}(0)$ and the isometric group of operators (\ref{grGs})
the following equality holds:
\begin{eqnarray}\label{idd}
&&\hskip-7mm\mathrm{Tr}_{s+1,\ldots,s+n}\,\prod_{Y_i\subset \mathrm{P}}\mathcal{G}^\ast_{|Y_{i}|}(t,Y_{i})
F_{s+n}(0,1,\ldots,s+n)=\mathrm{Tr}_{s+1,\ldots,s+n}\,F_{s+n}(0,1,\ldots,s+n).\nonumber
\end{eqnarray}
Then, taking into account for the subsets $Y\subset\,(s+1,\ldots,s+n)$ such equality
\begin{eqnarray*}
&&\sum\limits_{\mathrm{P}\,:(s+1,\ldots,s+n)\setminus Y={\bigcup\limits}_i Y_i}
(-1)^{|\mathrm{P}|}|\mathrm{P}|!=(-1)^{|(s+1,\ldots,s+n)\setminus Y|},
\end{eqnarray*} due to the validity for cumulant expansions of the expression (\ref{peregr}) for the solution of the BBGKY hierarchy,
we obtain the following image
\begin{eqnarray}\label{cherez1}
&&\hskip-4mm F_s(t,1,\ldots,s)=\\
&&\sum_{n=0}^\infty\frac{1}{n!}\,
\mathrm{Tr}_{s+1,\ldots,s+n}\,U_{1+n}^\ast(t,\{1,\ldots,s\},s+1,\ldots,s+n)F_{s+n}(0,1,\ldots,s+n),\quad s\geq1,\nonumber
\end{eqnarray}
where the symbol $U_{1+n}^\ast(t)$ denotes the $(1+n)$-order reduced cumulant of operator groups (\ref{grGs})
\begin{eqnarray*}
&&\hskip-4mm U_{1+n}^\ast(t,\{1,\ldots,s\},s+1,\ldots,s+n)=\\
&&\sum\limits_{Y\subset (s+1,\ldots,s+n)}(-1)^{|(s+1,\ldots,s+n)\setminus Y|}\,
\mathcal{G}^\ast_{|(1,\ldots,s)\cup Y|}(t,(1,\ldots,s)\cup Y).\nonumber
\end{eqnarray*}
Using the symmetry property of the initial reduced density operators, for the operators under the trace sign
in each term of the series (\ref{cherez1}) the following equalities hold:
\begin{eqnarray*}
&&\sum\limits_{Y\subset (s+1,\ldots,s+n)}
(-1)^{|(s+1,\ldots,s+n)\setminus Y)|}\,\mathcal{G}^\ast_{|(1,\ldots,s)\cup Y|}(t,(1,\ldots,s)\cup Y)
F_{s+n}(0)=\\
&&\sum\limits_{k=0}^{n}(-1)^{k}\sum\limits_{i_{1}<\ldots<i_{n-k}=s+1}^{s+n}
\mathcal{G}^\ast_{s+n-k}(t,1,\ldots,s,i_{1},\ldots,i_{n-k})F_{s+n}(0)=\\
&&\sum\limits_{k=0}^{n}(-1)^{k}\frac{n!}{k!(n-k)!}\,
\mathcal{G}^\ast_{s+n-k}(t,1,\ldots,s+n-k)F_{s+n}(0).\nonumber
\end{eqnarray*}
Thus, the reduced cumulant $(1+n)$-th order is represented by the following expansion \cite{L61}:
\begin{eqnarray*}
&&\hskip-7mm U_{1+n}^\ast(t,\{1,\ldots,s\},s+1,\ldots,s+n)=
\sum^n_{k=0}(-1)^k \frac{n!}{k!(n-k)!}\,\mathcal{G}_{s+n-k}^\ast(t,1,\ldots,s+n-k),
\end{eqnarray*}
and, as a result, we derive the representation for the expansions in the series of the BBGKY hierarchy, which in terms
of the operator (\ref{a}) takes the form of the following group of operators of the BBGKY hierarchy \cite{P95}:
\begin{eqnarray}\label{rcexp}
&& \hskip-7mm F(t)=\sum\limits_{n=0}^{\infty}\frac{1}{n!}\,\sum\limits_{k=0}^{n}\,(-1)^{k}\,
\frac{n!}{k!(n-k)!}\,\mathfrak{a}^{n-k}\mathcal{G}^\ast(t)\mathfrak{a}^{k}F(0)=\\
&& \mathcal{G}^\ast(t)F(0)+\sum\limits_{n=1}^{\infty}\frac{1}{n!}
\big[\underbrace{\mathfrak{a},\ldots,\big[{\mathfrak{a}}}_{\hbox{n-times}},
\mathcal{G}^\ast(t)\big]\ldots\big]F(0)=\nonumber\\
&& e^{\mathfrak{a}}\mathcal{G}^\ast(t)e^{-\mathfrak{a}}F(0).\nonumber
\end{eqnarray}

Finally, given the equality
\begin{eqnarray*}
&&\mathcal{G}^\ast(t-\tau)\big[\mathfrak{a},\mathcal{N}^\ast\big]\mathcal{G}^\ast(\tau)F(0)=
\frac{d}{d\tau}\mathcal{G}^\ast(t-\tau)\mathfrak{a}\mathcal{G}^\ast(t(\tau)F(0),
\end{eqnarray*}
the series expansion (\ref{rcexp}) is represented in the form of a perturbation theory series (series of iterations)
of the BBGKY hierarchy (\ref{h})
\begin{eqnarray*}
&&\hskip-8mm F(t)=\sum\limits_{n=0}^{\infty}\,\int\limits_{0}^{t} dt_{1}\ldots\int\limits_{0}^{t_{n-1}}dt_{n}
\mathcal{G}^\ast(t-t_{1})\big[\mathfrak{a},\mathcal{N}^\ast\big]\mathcal{G}^\ast(t_1-t_2)\ldots \mathcal{G}^\ast(t_{n-1}-t_n)\big[\mathfrak{a},\mathcal{N}^\ast\big]\mathcal{G}^\ast(t_{n})F(0),\nonumber
\end{eqnarray*}
or component by component \cite{Pe71}:
\begin{eqnarray*}\label{iter}
&&\hskip-7mm F_s(t,1,\ldots,s)=\sum\limits_{n=0}^{\infty}\,\int\limits_{0}^{t}dt_{1}\ldots\int\limits_{0}^{t_{n-1}}dt_{n}
\, \mathrm{Tr}_{s+1,\ldots,s+n}\,\mathcal{G}^\ast_s(t-t_{1})\sum\limits_{j_1=1}^{s}\mathcal{N}^{\ast}_{\mathrm{int}}(j_1,s+1)
\mathcal{G}^\ast_{s+1}(t_1-t_2)\ldots\\
&&\hskip+7mm \mathcal{G}^\ast_{s+n-1}(t_{n-1}-t_n)\sum\limits_{j_n=1}^{s+n-1}\mathcal{N}^{\ast}_{\mathrm{int}}(j_n,s+n)
\mathcal{G}^\ast_{s+n}(t_{n})F_{s+n}(0,1,\ldots,s+n), \quad s\geq1,\nonumber
\end{eqnarray*}
where the operator $\mathcal{N}^{\ast}_{\mathrm{int}}$ is determined by the particle interaction potential (\ref{h}).

\textcolor{blue!50!black}{\section{Nonperturbative solution of the hierarchy of evolution equations
for reduced observables}}
The motivation for describing the evolution of many-quantum particle systems using reduced observables (\ref{moo})
is related to possible equivalent representations of the mean value (expectation) functional of the observables (\ref{F(t)}).

The nonperturbative solution of the Cauchy problem for the hierarchy of evolution equations (\ref{dh}),(\ref{dhi})
for a sequence of reduced observables is represented by the following expansions \cite{GB}
\begin{eqnarray}\label{sed}
&&\hskip-8mm B_{s}(t,1,\ldots,s)=\sum_{n=0}^s\,\frac{1}{n!}\sum_{j_1\neq\ldots\neq j_{n}=1}^s
\mathfrak{A}_{1+n}(t,\{(1,\ldots,s)\setminus(j_1,\ldots,j_{n})\},\\
&&j_1,\ldots,j_{n}\big)\,
B_{s-n}(0,(1,\ldots,s)\setminus(j_1,\ldots,j_{n}))\otimes\mathbbm{1}_{(j_1)}\ldots\otimes\mathbbm{1}_{(j_n)},
\quad s\geq1,\nonumber
\end{eqnarray}
where the generating operators $\mathfrak{A}_{1+n}(t),\,n\geq0,$ are represented by cumulative expansions \eqref{cum}
over the operator groups (\ref{grG}) of quantum particle systems.

Cumulants of operator groups (\ref{grG}) of \eqref{sed} are defined as solutions of cluster expansions \eqref{exp}
of operator groups (\ref{grG}), namely by the following recursive relations:
\begin{eqnarray}\label{cexd}
&&\hskip-7mm \mathcal{G}_{s}(t,(1,\ldots,s)\setminus(j_1,\ldots,j_{n}),j_1,\ldots,j_{n})=\\
&&\sum\limits_{\mathrm{P}:\,(\{(1,\ldots,s)\setminus(j_1,\ldots,j_{n})\},\,j_1,\ldots,j_{n})=
\bigcup_i X_i}\,\prod\limits_{X_i\subset\mathrm{P}}\mathfrak{A}_{|X_i|}(t,X_i), \quad n\geq 0,\nonumber
\end{eqnarray}
where the symbol ${\sum}_{\mathrm{P}}$ denotes the sum over all possible partitions of $\mathrm{P}$ of the set of the
indeces $(\{(1,\ldots,s)\setminus(j_1,\ldots,j_{n})\},\,j_1,\ldots,j_{n})$ into $|\mathrm{P}|$ non-empty subsets
$X_i\subset(1,\ldots,s)$ that do not intersect. Therefore, the cumulant of the $(1+n)$th order of the operator groups
of a cluster of $s-n$ quantum particles and $n$ particles is given by the following expansions:
\begin{eqnarray}\label{cumulantd}
&&\hskip-5mm \mathfrak{A}_{1+n}(t,\{(1,\ldots,s)\setminus(j_1,\ldots,j_{n})\},j_1,\ldots,j_{n})\doteq\\
&&\sum\limits_{\mathrm{P}:\,(\{(1,\ldots,s)\setminus(j_1,\ldots,j_{n})\},\,j_1,\ldots,j_{n})={\bigcup}_i X_i}
(-1)^{\mathrm{|P|}-1}({\mathrm{|P|}-1})!\prod_{X_i\subset\mathrm{P}}
\mathcal{G}_{|\theta(X_i)|}(t,\theta(X_i)),\nonumber
\end{eqnarray}
where the symbol $\{(1,\ldots,s)\setminus(j_1,\ldots,j_{n})\}$ denotes a set consisting of one element
of indices, namely $(1,\ldots,s)\setminus(j_1,\ldots,j_{n})$, the declustering mapping $\theta$ is defined by
the formula: $\theta(\{X_i\})=(X_i)$ and the \eqref{cum} notations introduced above are used.

Examples of expansions (\ref{sed}) of reduced observables are given:
\begin{eqnarray*}
&&\hskip-7mm B_{1}(t,1)=\mathfrak{A}_{1}(t,1)B_{1}(0,1),\\
&&\hskip-7mm B_{2}(t,1,2)=\mathfrak{A}_{1}(t,\{1,2\})B_{2}(0,1,2)+
\mathfrak{A}_{2}(t,1,2)(B_{1}(0,1)\otimes\mathbbm{1}_{(2)}+\mathbbm{1}_{(1)}\otimes B_{1}(0,2)),\\
&&\hskip-7mm\vdots
\end{eqnarray*}

Note that the one-component sequences of reduced observables $B^{(1)}(0)=(0,b_{1}(1),0,\ldots)$
correspond to observables of additive type, and the one-component sequences of reduced observables
$B^{(k)}(0)=(0,\ldots,b_{k}(1,\ldots,k),0,\ldots)$ correspond to observables of $k$-ary type \cite{GB}.
If the initial condition (\ref{dhi}) is given by a sequence of reduced observables of additive type,
then the structure of the solution expansion (\ref{sed}) takes the form
\begin{eqnarray*}\label{af}
&&\hskip-7mm B_{s}^{(1)}(t,1,\ldots,s)=\mathfrak{A}_{s}(t,1,\ldots,s)
\sum_{j=1}^s \mathbbm{1}_{(1)}\otimes\ldots
\otimes\mathbbm{1}_{(j-1)}\otimes b_{1}(j)\otimes\mathbbm{1}_{(j+1)}\otimes\ldots
\otimes\mathbbm{1}_{(s)}, \quad s\geq 1,
\end{eqnarray*}
whose generating operator $\mathfrak{A}_{s}(t)$ is the cumulant of the $s$th order (\ref{cumulantd}) of groups
of operators (\ref{grG}).
In the case of a sequence of initial reduced observables of $k$-ary type, $k\geq2$, the solution schedules
(\ref{sed}) take the following form:
\begin{eqnarray*}\label{af-k}
&&\hskip-5mm B_{s}^{(k)}(t)=0, \quad 1\leq s<k,\\
&&\hskip-5mm B_{s}^{(k)}(t,1,\ldots,s)=\frac{1}{(s-k)!}\sum_{j_1\neq\ldots\neq j_{s-k}=1}^s
\mathfrak{A}_{1+s-k}\big(t,\{(1,\ldots,s)\setminus (j_1,\ldots,j_{s-k})\},\nonumber\\
&& j_1,\ldots,j_{s-k}\big)b_{k}((1,\ldots,s)\setminus(j_1,\ldots,j_{s-k}))
\otimes\mathbbm{1}_{(j_1)}\otimes \ldots\otimes\mathbbm{1}_{(j_{s-k})},\quad s\geq k,\nonumber
\end{eqnarray*}
where the generating operator $\mathfrak{A}_{1+s-k}(t)$ is the cumulant (\ref{cumulantd}) of the $(1+s-k)$th order.

If $b_{s}\in\mathfrak{L}(\mathcal{H}_{s})$, then according to the estimate \eqref{est} for the generating operator
(\ref{cumulantd}) of the expansion for reduced observables \eqref{sed} we have
\begin{eqnarray*}
&&\hskip-5mm \big\|\mathfrak{A}_{1+n}(t)b_{s}\big\|_{\mathfrak{L}(\mathcal{H}_{s})}
\leq n!\,e^{n+2}\,\big\|b_{s}\big\|_{\mathfrak{L}(\mathcal{H}_{s})}.
\end{eqnarray*}
Therefore, under the condition $\gamma<e^{-1}$ for a sequence of reduced observables (\ref{se}) in the space
$\mathfrak{L}_{\gamma}(\mathcal{F}_{\mathcal{H}})$ the following inequality holds
\begin{eqnarray*}\label{Fes}
&&\|B(t)\|_{\mathfrak{L}_{\gamma}(\mathcal{F}_{\mathcal{H}})}\leq
e^2\,(1-\gamma e)^{-1}\,\|B(0)\|_{\mathfrak{L}_{\gamma}(\mathcal{F}_{\mathcal{H}})},
\end{eqnarray*}
and the functional \eqref{B(t)} exists.

The following criterion holds.
\begin{criterion} A sequence of reduced observables represented by expansions (\ref{sed})
if and only if is a solution of the Cauchy problem for a hierarchy of evolution equations (\ref{dh}),(\ref{dhi}),
if the generating operators (\ref{cumulantd}) of the decompositions (\ref{sed}) are solutions of the cluster expansions (\ref{cexd})
of the operator groups (\ref{grG}) of the Heisenberg equations.
\end{criterion}

The necessity condition means that for the operator groups (\ref{grG}) there are cluster expansions (\ref{cexd}).
These recursive relations are derived from the definition (\ref{moo}) of reduced observables provided that they are represented
by expansions (\ref{sed}) for the solution of the Cauchy problem of the hierarchy of evolution equations (\ref{dh}),(\ref{dhi}).

A sufficient condition means that the infinitesimal generator of the one-parameter mapping (\ref{sed}) coincides
with the generator of the sequence of recurrent evolution equations (\ref{dh}). Indeed, in the space
$\mathfrak{L}_{\gamma}(\mathcal{F}_{\mathcal{H}})$ the following existence theorem holds \cite{GB}.
\begin{theorem}
The nonperturbative solution of the Cauchy problem (\ref{dh}),(\ref{dhi}) is represented by expansions (\ref{sed}),
in which the generating operators are the cumulants (\ref{cumulantd}) of the groups of operators (\ref{grG}) of the corresponding order).

Under the condition $\gamma<e^{-1}$ for the initial data $B(0)\in \mathfrak{L}_{0}(\mathcal{F}_{\mathcal{H}})$ the sequence
(\ref{sed}) is the unique global classical solution, and for arbitrary initial data
$B(0)\in \mathfrak{L}_{\gamma}(\mathcal{F}_{\mathcal{H}})$ the unique global generalized solution.
\end{theorem}

The proof of the theorem is analogous to the proof of the existence theorem for the Heisenberg equations (\ref{A(t)}) of many
quantum particles in the space of sequences of bounded operators \cite{GB}. A similar result is also valid for the sequences
of unbounded operators observed from the space of sequences \cite{Ger12}.

\textcolor{blue!50!black}{\section{Representation of reduced observables by perturbation theory}}
Cluster expansions (\ref{cexd}) of groups of operators (\ref{grG}) form the basis for the classification of all possible
representations of solutions to the Cauchy problem for the hierarchy of evolution equations (\ref{dh}),(\ref{dhi}) for
reduced observables.

Indeed, using cluster expansions (\ref{cexd}) of operator groups (\ref{grG}), one can construct other representations of solutions.
For example, let us express the cumulants $\mathfrak{A}_{1+n}(t),\,n\geq2,$ of operator groups
(\ref{grG}) in terms of cumulants of $1$ and $2$ order. The following equalities hold:
\begin{eqnarray*}
&&\hskip-5mm\mathfrak{A}_{1+n}(t,\{(1,\ldots,s)\setminus(j_1,\ldots,j_{n})\},j_1,\ldots,j_{n})= \\
&&\sum_{Y\subset(j_1,\ldots,j_{n}),\, Y\neq \emptyset}
\mathfrak{A}_{2}(t,\{(1,\ldots,s)\setminus(j_1,\ldots,j_{n})\},\{Y\})
\sum_{\mathrm{P}:\,(j_1,\ldots,j_{n})\setminus Y ={\bigcup\limits}_i X_i} (-1)^{|\mathrm{P}|}\,|\mathrm{P}|!\,
\prod_{i=1}^{|\mathrm{P}|}\mathfrak{A}_{1}(t,\{X_{i}\}),\\
&&\hskip-5mm n\geq2,
\end{eqnarray*}
where the symbol ${\sum\limits}_{Y\subset(j_1,\ldots,j_{n}),\,Y\neq\emptyset}$ denotes the sum over all non-empty
subsets of the indices $Y\subset (j_1,\ldots,j_{n})$.

According to identity
\begin{eqnarray}\label{id}
&&\hskip-7mm\sum_{\mathrm{P}:\,(j_1,\ldots,j_{n})\setminus Y ={\bigcup\limits}_i X_i}
(-1)^{|\mathrm{P}|}\,|\mathrm{P}|!\,\prod_{i=1}^{|\mathrm{P}|}
\mathfrak{A}_{1}(t,\{X_{i}\})B_{s-n}(0,(1,\ldots,s)\setminus(j_1,\ldots,j_{n}))\otimes\mathbbm{1}_{(j_1)}\ldots\otimes\mathbbm{1}_{(j_{n})}=
\nonumber\\
&&\hskip-5mm\sum_{\mathrm{P}:\,(j_1,\ldots,j_{n})\setminus Y ={\bigcup\limits}_i X_i}
(-1)^{|\mathrm{P}|}\,|\mathrm{P}|!\,B_{s-n}(0,(1,\ldots,s)\setminus(j_1,\ldots,j_{n}))
\otimes\mathbbm{1}_{(j_1)}\ldots\otimes\mathbbm{1}_{(j_{n})}, \nonumber
\end{eqnarray}
and for $Y\subset\,(j_1,\ldots,j_{n})$ equalities
\begin{eqnarray}\label{aq}
&&\hskip-7mm {\sum_{\mathrm{P}:\,(j_1,\ldots,j_{n})\setminus Y =
{\bigcup\limits}_i X_i}}(-1)^{|\mathrm{P}|}\,|\mathrm{P}|!=
(-1)^{|(j_1,\ldots,j_{n})\setminus Y|},
\end{eqnarray}
for solution schedules (\ref{sed}), we output the following image:
\begin{eqnarray*}
&&\hskip-5mm B_{s}(t,1,\ldots,s)=\sum_{n=1}^s\,\frac{1}{n!}\,\sum_{j_1\neq\ldots\neq j_{n}=1}^s\,\,
\sum\limits_{Y\subset (1,\ldots,s)\setminus(j_1,\ldots,j_{n}),\,Y\neq \emptyset}\,
(-1)^{|(j_1,\ldots,j_{n})\setminus Y|}\times\\
&&\mathfrak{A}_{2}(t,\{j_1,\ldots,j_{n}\},\{Y\})\,B_{s-n}(0,(1,\ldots,s)\setminus(j_1,\ldots,j_{n}))
\otimes\mathbbm{1}_{(j_1)}\ldots\otimes\mathbbm{1}_{(j_{n})}, \quad s\geq 1,
\end{eqnarray*}
where the above accepted notations are used.

An analogue of the representation of the solution of the BBGKY hierarchy for quantum systems, constructed in the works
for classical particle systems \cite{GG21}, in the case of the hierarchy of evolution equations for reduced observables,
is represented by expansions with generating operators, which are reduced cumulants of groups of operators of quantum
systems of a finite number of particles.
Taking into account the validity of the identity (\ref{id}) for the initial reduced observables, we derive the reduced
representation of the distributions (\ref{sed}):
\begin{eqnarray}\label{rsedc}
&&\hskip-7mm B(t)=\sum\limits_{n=0}^{\infty}\frac{1}{n!}\,\sum\limits_{k=0}^{n}\,(-1)^{n-k}\,
\frac{n!}{k!(n-k)!}\,(\mathfrak{a}^{+})^{n-k}\mathcal{G}(t)(\mathfrak{a}^{+})^{k}B(0)=\nonumber\\
&&\mathcal{G}(t)B(0)+\sum\limits_{n=1}^{\infty}\frac{1}{n!}
\big[\ldots\big[\mathcal{G}(t),\underbrace{\mathfrak{a}^{+}\big],\ldots,\mathfrak{a}^{+}}_{\hbox{n-times}}\big]B(0)=\nonumber\\
&&e^{-\mathfrak{a}^{+}}\mathcal{G}(t)e^{\mathfrak{a}^{+}}B(0).\nonumber
\end{eqnarray}
Therefore, in component-wise form, the generating operators of the expansions of the solution depicted in expansion form (\ref{sed})
are the following
reduced cumulants of groups of operators (\ref{grG}):
\begin{eqnarray}\label{rcc}
&&\hskip-7mm U_{1+n}(t,\{1,\dots,s-n\},s-n+1,\dots,s)=\\
&&\sum^n_{k=0}(-1)^{k}\frac{n!}{k!(n-k)!}\mathcal{G}_{s-k}(t,1,\dots,s-k)\otimes\mathbbm{1}_{(s-k+1)}\otimes\ldots
\otimes\mathbbm{1}_{(s)},
\quad n\geq 0. \nonumber
\end{eqnarray}
Indeed, the solutions of the recursive relations (\ref{cexd}) with respect to the first-order cumulants can be
represented as expansions over the cumulants acting in the subspaces in which the initial reduced observables act,
and over the cumulants acting in other subspaces.
\begin{eqnarray*}
&&\hskip-7mm \mathfrak{A}_{1+n}(t,\{(1,\ldots,s)\setminus(j_1,\ldots,j_{n})\},j_1,\ldots,j_{n})=\\
&&\sum_{Y\subset (j_1,\ldots,j_{n})}\mathfrak{A}_{1}(t,\{(1,\ldots,s)\setminus((j_1,\ldots,j_{n})\cup Y)\})
\sum_{\mathrm{P}:\,(j_1,\ldots,j_{n})\setminus Y ={\bigcup\limits}_i X_i}
(-1)^{|\mathrm{P}|}\,|\mathrm{P}|!\,\prod_{i=1}^{|\mathrm{P}|}\mathfrak{A}_{1}(t,\{X_{i}\}),
\end{eqnarray*}
where the symbol ${\sum\limits}_{Y\subset(j_1,\ldots,j_{n})}$ denotes the sum over all possible subsets
$Y\subset(j_1,\ldots,j_{n})$. Then, taking into account the identity (\ref{id}) and equalities (\ref{aq}),
we derive the expansions (\ref{rsedc}), the generating operators of which are reduced cumulants (\ref{rcc}).

As noted, traditionally, the solution of the BBGKY hierarchy for the states of many quantum particles is represented
by a perturbation theory series \cite{P95}. The expansions (\ref{rsedc}) can also be represented as expansions of
perturbation theory \cite{GB}:
\begin{eqnarray*}
&& \hskip-7mm B(t)=\sum\limits_{n=0}^{\infty}\,\int\limits_{0}^{t} dt_{1}\ldots
\int\limits_{0}^{t_{n-1}}dt_{n}\,\mathcal{G}(t-t_{1})\big[\mathcal{N},\mathfrak{a}^+\big]
\mathcal{G}(t_1-t_2)\ldots \mathcal{G}(t_{n-1}-t_n)\big[\mathcal{N},\mathfrak{a}^+\big]\mathcal{G}(t_{n})B(0).
\end{eqnarray*}
Indeed, as a result of applying analogs of the Duhamel equation to generating operators (\ref{cumulantd})
of expansions (\ref{sed}) in component-wise form, we obtain, for example,
\begin{eqnarray*}
&& \hskip-5mm U_{1}(t,\{1,\ldots,s\})=\mathcal{G}_s(t,1,\ldots,s),\\
&& \hskip-5mm U_{2}(t,\{(1,\ldots,s)\setminus(j_1)\},j_1)=\\
&&\int\limits_{0}^{t}dt_{1}\,\mathcal{G}_s(t-t_{1},1,\ldots,s)
\sum_{j_{2}=1,\,j_2\neq j_{1}}^s\mathcal{N}_{\mathrm{int}}(j_1,j_{2})\,
\mathcal{G}_{s-1}(t_1,(1,\ldots,s)\setminus j_1)\otimes\mathbbm{1}_{(j_{1})},\\
&&\hskip-2mm\vdots
\end{eqnarray*}
where the operators $\mathcal{N}_{\mathrm{int}}(j_1,j_{2})$ are determined by the pair interaction potential
of the particles (\ref{dh}).

\textcolor{blue!50!black}{\section{Conclusion}}
As is well known \cite{Ger12}, due to the validity of the equality (\ref{F(t)}) for the representations
of the mean-value functional of the observables, there are two equivalent approaches to describing the
evolution of quantum systems of many particles. Specifically, one can describe the evolution in terms
of observables, which is governed by a hierarchy of evolution equations (\ref{dh}), or in terms of a
state, where the evolution is described by the BBGKY hierarchy (\ref{h}). For systems with a finite
number of particles, these hierarchies of evolution equations are equivalent to the Heisenberg equation
\eqref{A(t)} and the von Neumann equation \eqref{D(t)}, respectively.

The article substantiates the structure of expansions that represent the nonperturbative solution of the
Cauchy problem for hierarchies of evolution equations for a sequence of reduced observable systems of
many quantum particles, as well as expansions into series that represent the nonperturbative solution of
the Cauchy problem for the BBGKY hierarchy for a sequence of reduced density operators that describe the
evolution of the state of quantum systems. It is established that the classification of representations
of solutions to the Cauchy problem for hierarchies of evolution equations for the state (\ref{h}) and
observable (\ref{dh}) of many-particle quantum systems is based on the cluster expansions (\ref{cex})
of operator groups (\ref{grGs}) for the sequence of von Neumann equations (\ref{D(t)}) and the cluster
expansions (\ref{cexd}) of operator groups (\ref{grG}) of the Heisenberg equations (\ref{A(t)}), respectively.

The structure of the nonperturbative solution of the BBGKY hierarchy (\ref{h}) has the form of a series
expansion in particle groups (\ref{se}), the generating operators of which are determined by the cumulants
(\ref{cumulant}) of the corresponding order for the particle cluster and the particles of the adjoint
operator groups (\ref{grGs}). The nonperturbative solution of the recurrent evolution equations for reduced
observables (\ref{dh}) of quantum systems is represented by a particle group expansion (\ref{sed}), the
generating operators of which are determined by the cumulants (\ref{cumulantd}) of the corresponding order
for the particle cluster and the particles of the operator groups (\ref{grG}). The generating operators of
the solutions of the hierarchies of evolutionary equations (\ref{h}) and (\ref{dh}) of many-particle quantum
systems are dual in the sense of the mean value functional \eqref{averageD}.

Initial states from the space of sequences of trace-class operators
$\mathfrak{L}^{1}_\alpha(\mathcal{F}_{\mathcal{H}})$ describe systems of a finite average number of particles
\cite{Ger12}. For such initial states, due to the validity of the equalities (\ref{idd}), all possible
representations of the generating operators of the solution of the BBGKY hierarchy (\ref{h}) are equivalent.
For the states of quantum systems of an infinite number of particles, such representations are not equivalent,
and the solution is represented by series expansions (\ref{se}). For initial observables from the space
$\mathfrak{L}_\gamma(\mathcal{F}_{\mathcal{H}})$ in the case of a hierarchy of evolution equations for reduced
observables (\ref{dh}) due to the validity of the identities (\ref{id}) all possible representations of the
generating operators of the solution (\ref{sed}) are equivalent.

Note that the work considered quantum systems of an unspecified number of identical spinless particles,
satisfying the Maxwell--Boltzmann statistics. The results formulated above can be extended to systems of
many bosons or fermions in accordance with the work \cite{GP}.

Recently, in the articles \cite{GT},\cite{GTrmp} (in terms of observables \cite{G11},\cite{G15}) using
an analogue of the method of cluster expansions of operator groups, the so-called kinetic cluster expansions
of cumulants of operator groups of quantum systems of many particles, a generalized quantum kinetic equation
was rigorously derived and a sequence of state functionals was constructed, which describe the evolution of
correlations of quantum systems using such a kinetic equation (see also the review \cite{GG21} for classical
particle systems with collisions).

Thus, the construction of nonperturbative solutions to hierarchies of fundamental evolution equations is based
on the concept of cumulants of operator groups and also underlies the kinetic description of the collective
behavior of quantum systems of an infinite number of particles.

\vskip+5mm

\noindent \textbf{Acknowledgements.} \,\, The work was supported by
“Research during the war in the largest Ukrainian mathematical institution, Grant No. SFIPD-Ukraine-00014586”

\medskip


\addcontentsline{toc}{section}{\textcolor{blue!55!black}{References}}

\vskip+5mm

\end{document}